# Spin-biased Quantum Spin Hall Effect in Altermagnetic Lieb Lattice

Qianjun Wang[a,b], Ruqian Wu[c,*]   Jun Hu[b,†]

[a] School of Physical Science and Technology, Soochow University, Suzhou 215006, China

[b] Institute of High Pressure Physics, School of Physical Science and Technology, Ningbo University, Ningbo 315211, China

[c] Department of Physics and Astronomy, University of California, Irvine, California 92697, USA

Altermagnetic (AM) order, a recently discovered magnetic state, has attracted intense research interest for its potential applications in spintronic and quantum technologies. Here, we theoretically investigate the AM state in the Lieb lattice, a prototypical two-dimensional lattice, using the Hubbard model. We show that AM order emerges with only moderate electronic correlations. Strikingly, spin-orbit coupling drives the system into a topological phase exhibiting a new quantum spin Hall effect (QSHE) with spin-biased topological edge states in one-dimensional nanoribbons. These edge states possess different localizations and velocities, and hence may produce spin and charge currents, fundamentally distinct from that in conventional topological insulators with spin degeneracy. This novel spin-biased QSHE in the AM Lieb lattice unveils exciting opportunities for both fundamental studies and innovative device concepts, motivating immediate experimental exploration.

[*] Email: wur@uci.edu
[†] Email: hujun2@nbu.edu.cn

Altermagnetism represents a new symmetry distinct class of magnetic order, arising from an expanded classification of antiferromagnetic (AFM) materials that were traditionally characterized solely by their compensated spin moments. In altermagnetic (AM) order, however, the two opposite magnetic sublattices are connected by rotation or mirror symmetries, rather than by inversion or translation symmetries as in conventional collinear AFM materials. This key symmetry distinction gives rise to non-relativistic momentum-dependent spin splitting, a feature typically associated with ferromagnetic (FM) materials[1,2]. As a result, altermagnets simultaneously break time-reversal symmetry (TRS) and maintain zero net magnetization, combining advantages of both FM and AFM states in a single phase. This dual character has attracted great fundamental interest in condensed matter physics and materials science, as it supports robust spin polarization without stray fields and opens pathways to low-dissipation and high-coherence spin transport for next-generation spintronic applications[3,4,5,6]. Consequently, extensive efforts have focused on exploring realistic AM materials. So far, several materials, including $MnTe$[7,8,9], $CrSb$[10,11,12], and $Mn_5Si_3$[13], have been experimentally confirmed as altermagnets, while the proposed AM order in $RuO_2$ remains under active debate[14,15]. More recently, the altermagnetism framework has been generalized to include non-colinear AFM materials, further expanding the landscape of candidate materials and broadening the range of accessible spintronic functionalities[16,17,18].

In parallel, researchers have also explored the possibility of realizing AM order in two-dimensional (2D) systems through theoretical models and first-principles calculations, motivated by the tunability and integrability of 2D materials for development of nanoscale electronic and spintronic devices[19,20]. Although the pristine honeycomb lattice is centrosymmetric and thus prohibits the existence of AM order, a modified honeycomb lattice with specific dimerization of nearest-neighboring spins may host AM order and exhibit topological features, such as Dirac or Weyl states and quantum spin or anomalous Hall effects, as predicted via tight-binding models[21,22]. First-principles calculations have further demonstrated that $MnPX_3$ (X = S, Se), which hosts a honeycomb Mn sublattice, can be transformed into an altermagnet when inversion symmetry is lifted, for example, through Janus structures or by interfacing with suitable substrates[23,24,25]. Additionally, twisted van der Waals (vdW) bilayers of honeycomb-lattice-based materials may also develop AM order due to twist-induced centrosymmetry breaking[26,27,28,29], with certain systems predicted to exhibit the quantum spin Hall effect (QSHE)[28].

Furthermore, the pristine Lieb lattice provides an ideal platform for realizing AM order when its B and C sublattices carry opposite spins [see Fig. 1(a)], as it naturally satisfies the symmetry criteria required for altermagnetism[30,31,32,33]. Indeed, several 2D vdW materials predicted to be altermagnets can be viewed as derivatives of the Lieb lattice, many of which also exhibit notable topological characteristics[33,34,35,36,37,38,39]. However, the experimental realization of genuinely 2D Lieb-lattice-based altermagnets has remained elusive. Interestingly, the Lieb lattice serves as the key structural motif in the quasi-2D altermagnets $KV_2Se_2O$[40], $RbV_2Te_2O$[41], $Cs_{1-\delta}V_2Te_2O$[42], and $La_2O_3Mn_2Se_2$[43], suggesting its strong potential for realizing 2D AM phases. Prior theoretical work based on the Hubbard model primarily focused on the regime of strong electronic correlations[32]. The nature of AM order under weak and moderate electronic correlations and the possible emergence of topological features remains largely unexplored.

In this Letter, we use the Hubbard model to investigate the electronic and magnetic properties of the Lieb lattice in the regime of weak-to-moderate electronic correlations. By tuning the onsite chemical potentials and interaction strength, we identify three distinct phases: a nonmagnetic (NM) metal, an AM metal and an AM insulator. Intriguingly, the inclusion of spin-orbit coupling (SOC) drives the AM metal into an insulating state, opening a band gap that scales linearly with the SOC strength. Moreover, for AM insulators with small intrinsic gaps, SOC induces a band inversion. Both categories of SOC-driven insulating states are topological insulators (TIs) that host QSHE, representing a new class of topological AM materials.

We start by analyzing the Hamiltonian of the Lieb-lattice Hubbard model, which takes the form

$$\widehat{H} = \widehat{H}_0 + \widehat{H}_U + \widehat{H}_{SO}. \quad (1)$$

The first term $\widehat{H}_0$ describes the nearest- and next-nearest-neighbor hopping as well as the onsite energy:

$$\widehat{H}_0 = t_1 \sum_{\langle i,j \rangle,\sigma} c_{i\sigma}^\dagger c_{j\sigma} + t_2 \sum_{\langle\langle i,j \rangle\rangle,\sigma} c_{i\sigma}^\dagger c_{j\sigma} + \frac{\Delta}{2} \sum_{i \in A,\sigma} n_{i\sigma} - \frac{\Delta}{2} \sum_{i \in B,C,\sigma} n_{i\sigma}. \quad (2)$$

Here, $c_{i\sigma}^\dagger$ ($c_{i\sigma}$) creates (annihilates) an electron with spin $\sigma$ at site $i$, and $n_{i\sigma}$ denotes the electron number operator. The parameters $t_1$ and $t_2$ stand for the nearest- and next-nearest-neighbor hopping amplitudes, respectively. We set $t_1 = -t$ and $t_2 = -0.1t$, where $t > 0$ defines the magnitude of the nearest-neighbor hopping energy and serves as the energy unit throughout this work. The parameter $\Delta$ measures the onsite potential difference between the A and B/C sublattices, with $\Delta > 0$ indicating that electrons preferentially occupy the B and C sites. In the AM order, the B and C sublattices carry opposite spins, which naturally breaks the fourfold symmetry of each spin channel.

The second term, $\hat{H}_U$, represents the local onsite Coulomb repulsion $U$ acting on the B and C sublattices[32]:

$$\hat{H}_U = U \sum_{i \in B,C} n_{i,\uparrow} n_{i,\downarrow}. \tag{3}$$

A second-order perturbation expansion in $t/U$ yields an effective Heisenberg interaction with exchange $J = \frac{4t^2}{U} > 0$. This superexchange interaction energetically favors antiferromagnetic alignment between the B and C sublattices, which is essential for establishing the altermagnetic order because these sites are related by rotation and mirror symmetries of the Lieb lattice.

The third term, $\hat{H}_{SO}$, accounts for the SOC effect. In two-dimensional systems, SOC can generally be decomposed into intrinsic and Rashba components[44]. For systems possessing out-of-plane mirror symmetry with respect to the *xy*-plane, the Rashba term vanishes. As a result, the z-component of spin ($S_z$) is conserved, and $\hat{H}_{SO}$ does not induce spin-flip processes between the spin-up and spin-down channels. Moreover, SOC arises only between the B and C sublattices, because the A sites serve as centers of the rotational $C_{4v}$ symmetry. Under these symmetry constraints, $\hat{H}_{SO}$ in the Lieb lattice can therefore be expressed in the form of the Kane-Mele model[44]:

$$\hat{H}_{SO} = i\lambda \sum_{\langle\langle i\alpha, j\beta \rangle\rangle} (\boldsymbol{d}_i \times \boldsymbol{d}_j)_z \sigma^z_{\alpha\beta} c^\dagger_{i\alpha} c_{j\beta}, \tag{4}$$

where $\lambda$ denotes the SOC strength, $\sigma^z$ is the Pauli matrix, and $\boldsymbol{d}_i$ is the unit vector pointing from the A site toward the B or C site.

Since the emergence of AM order does not rely on SOC, we first set $\lambda = 0$ and focus on electron filling factor $n = 2$ under weak and moderate correlations with $U < 5t$. Figure 1(b) displays the phase diagram of the electronic and magnetic states with varying $\Delta$ and $U$. Three distinct phases are identified: (I) an NM metal, (II) an AM metal, and (III) an AM insulator. For weak correlation with $U < 0.7t$, the system remains an NM metal over the whole range of $\Delta$. As $U$ increases, the AM order gradually develops, and smaller values of $\Delta$ are required to stabilize it. The phase boundary between the NM metal and AM metal can be approximately fitted by $(\Delta - 0.5)(U - 0.3) = 1.8$. The AM metallic phase persists up to the boundary described by $(\Delta - 0.4)(U - 0.1) = 6.1$, beyond which the system transits into the AM insulating phase. Representative band structures of these phases are shown in Fig. 1(c). In the AM phases, the bands along the Γ-M direction preserve TRS, whereas pronounced spin splitting appears along other paths in the Brillouin zone. Additionally, the bands near the Fermi level ($E_F$) along the Γ-X-M path belong to spin-up electrons, while those along the Γ-Y-M path belong to spin-down electrons.

Both the local spin moment ($M_S$) and electron occupancy ($n_e$) depend on $\Delta$ and $U$, as shown in Fig. 1(d). For all selected values of $\Delta$, $M_S$ on either the B or C sublattice remains zero at small $U$, corresponding to the NM metallic phase. Once $U$ exceeds a critical threshold, which is larger for smaller $\Delta$, the magnitude of $M_S$ rapidly increases and then gradually saturates, indicating the onset of the AM phases. In contrast, $n_e$ on the B/C site initially decreases and then increases with increasing $U$ at fixed $\Delta$. This behavior originates from electron transfer from the B/C site to the A site driven by onsite Coulomb repulsion, which becomes more significant for small $\Delta$. For example, $n_e$ on the A site reaches 0.65 for $\Delta = 2t$ and $U = 2.7t$. In the AM regime, however, spin polarization mitigates the onsite Coulomb repulsion on the B and C sites, while the onsite potential difference $\Delta$ promotes electrons hopping from the A site back to the B and C sites. These cooperative effects reduce the total energy, leading to an increase in $n_e$ on the B and C sites with increasing $U$.

We next examine the influence of SOC on electronic properties. The band structures of the NM metallic and AM insulating phases are only weakly affected, whereas the AM metallic phase experiences significant modifications. As shown in Fig. 2(a), SOC opens a clear band gap at the crossing points near $E_F$ along the X-M and Y-M directions, in contrast to the gapless dispersion in the absence of SOC [Fig. 1(c)]. Moreover, SOC generates a global band gap across the entire Brillouin zone, as evidenced by the three-dimensional band structures of the spin-up and spin-down channels in Fig. 2(c) and 2(d). Interestingly, SOC opens additional indirect gaps in opposite momentum directions for both spin channels, effectively transforming the AM metal into an AM insulator. The valence-band maximum (VBM) is located at the M point and the conduction-band minimum (CBM) is at the X (or Y) point. The global band gap increases monotonically with SOC strength, as illustrated by the blue curve in Fig. 2(b). In addition, the local SOC-induced gaps are significantly larger and more sensitive to $\lambda$, reaching $0.29t$ at $\lambda = 0.05t$, as indicated by the red curve in Fig. 2(b). It is important to note that this SOC-driven AM insulating phase differs fundamentally from phase (III) in Fig. 1(b), which corresponds to a normal band insulator.

This observation leads to an intriguing question: is the SOC-induced AM state a TI? To answer this, we calculated the Berry curvature $\Omega(\mathbf{k})$ and Chern number $(C)$[45,46]:

$$\Omega(\mathbf{k}) = -2\,\text{Im} \sum_{n\in\{o\}} \sum_{m\in\{u\}} \frac{\langle\psi_{n\mathbf{k}}|v_x|\psi_{m\mathbf{k}}\rangle\langle\psi_{m\mathbf{k}}|v_y|\psi_{n\mathbf{k}}\rangle}{(\varepsilon_{m\mathbf{k}} - \varepsilon_{n\mathbf{k}})^2}, \quad (5)$$

and

$$C = \frac{1}{2\pi} \int_{BZ} \Omega(\mathbf{k})\,d\mathbf{k}, \quad (6)$$

Here, {o} and {u} denoting the sets of occupied and unoccupied states, respectively. $\psi_{n\mathbf{k}}$ and $\varepsilon_{n\mathbf{k}}$ are the Bloch wave function and eigenvalue of the $n$-th state at wave vector $\mathbf{k}$, and $v_{x(y)}$ is the velocity operator. The anomalous Hall conductance is expressed as $\sigma_{xy} = Ce^2/h$. Because of the conservation of $S_z$ in our model, the spin-up and spin-down channels are decoupled. Therefore, the integration in Eq. (6) can be carried out separately for the two spin channels, resulting in spin-resolved Chern numbers $C_+$ for the spin-up channel and $C_-$ for the spin-down channel, respectively. Accordingly, the total Chern number can be written as the sum of the two spin-resolved contributions, $C = C_+ + C_-$, and the same additive relation holds for the Hall conductance[47].

Figure 3(a) displays the spin-resolved anomalous Hall conductance $\sigma_{xy,\pm}$ for $\Delta = 2t$, $U = 3t$, and $\lambda = 0.01t$, $0.03t$, and $0.05t$. Remarkably, quantized plateaus with $C_\pm = \pm 1$ appear near $E_F$ in both spin channels, signifying the emergence of a topological phase. However, since $C = 0$, this topological state differs from the typical quantum anomalous Hall effect (QAHE). Instead, the system exhibits a nonzero integer spin Chern number: $C_S = (C_+ - C_-)/2 = 1$, corresponding to a quantized spin Hall conductance $\sigma_{xy}^S = C_S e^2/h = (\sigma_{xy,+} - \sigma_{xy,-})/2$, which manifests this phase as a QSHE state[44,48]. Notably, since the TRS is intrinsically broken is the AM order, the QSHE realized here represents a novel topological phase, distinct from that observed in nonmagnetic TIs, where TRS must be broken by moderate external perturbations such as magnetic fields or impurities[48,49,50]. Furthermore, the quantized values of $C_\pm$, and hence $\sigma_{xy}^S$, remain robust even when moderate perturbation to the out-of-plane mirror symmetry induces a finite Rashba SOC, similar to the behavior observed in graphene[51]. A systematic investigation of this robustness will be reported elsewhere.

Note that symmetry plays a crucial role in realizing quantized $\sigma_{xy,\pm}$ in the Lieb lattice. Structurally, the B and C sublattices are connected by $C_{4v}$ and mirror symmetries, resulting in identical band dispersions along the $k_x$ and $k_y$ directions, as shown in Fig. 2(c) and 2(d). However, opposite spins on the B and C sublattices reduce the symmetry to $C_{2v}$ in each spin channel and hence all bands are spin-polarized in the entire momentum space except along the diagonal Γ-M directions. The Berry curvature distributions in Fig. 3(b) and 3(c) reflect the symmetry reduction and the origin of opposite Hall conductivities from the two spin channels. Accordingly, topologically protected edge states are expected in one-dimensional (1D) nanoribbons[44], especially by cutting the Lieb lattice along the A-B or A-C directions. To verify this, we constructed a nanoribbon with a width of 20.5 unit cells, periodic direction along the A-C chains, and both edges terminated by an A-C chain [see the inset of Fig. 3(e)]. The corresponding band structure for $\Delta = 2t$, $U = 3t$, and $\lambda = 0.05t$ is shown in Fig. 3(d). Indeed, two bands cross $E_F$ around the Y point, and the spatial distributions of their wavefunctions in Fig. 3(e) indicate that they are strongly localized on edges.

For the same direction of propagation, different spin states take opposite edges [Fig. 3(e)], a characteristic feature of the QSHE, even though the underlying mechanism is different from that in nonmagnetic TIs. In conventional nonmagnetic TIs, spin-up and spin-down edge states on the same edge are degenerate in energy and possess opposite wave vectors, resulting in spin-unbiased edge transport. In contrast, the AM Lieb lattice hosts spin-biased edge states. Firstly, the spin-up and spin-down edge states are nondegenerate in energy. Secondly, the wave vectors of the spin-up and spin-down edge states on the same edge differ both in amplitude and direction. Thirdly, the spin-up edge state is more strongly localized. As a result, this spin-biased QSHE generates not only spin currents but also charge currents along the edges of the corresponding 1D nanoribbons. Accordingly, the resulting topological phase can be classified as AM QSHE, consistent with recent report on altermagnets with $C_{4z}T$ symmetry[52]. We emphasize that the emergence of the QSHE does not rely on a particular crystal structure; rather, it is governed by SOC-induced band inversion that opens a topological gap. However, the characteristics of the edge states are closely tied to the underlying lattice symmetry. For example, the AM QSHE reported in Lieb-lattice-like systems, such as the $Nb_2SeTeO$ monolayer and $Fe_2Se_2O$ multilayers, exhibits nondegenerate edge-state bands in the corresponding 1D nanoribbons[36,53], consistent with the spin-biased QSHE predicted in our Lieb-lattice model. In contrast, quantum spin Hall phases in honeycomb-lattice-based AM systems preserve symmetries that protect Kramers degeneracy at the edges, resulting in degenerate and spin-unbiased edge states that resemble those in nonmagnetic TIs[28,54]. These similarities and distinctions therefore originate from the specific symmetry properties of the honeycomb and Lieb lattices.

Another interesting phenomenon is observed as we further increase U across the phase boundary between the AM metal and AM insulator. In the AM metallic phase without SOC, two bands cross $E_F$ between the M and X (or Y) points, leaving a gap between the $X_2$ and $X_3$ energy levels [Fig. 1(c)]. As the value of $U$ increases, these intersecting points move toward the X or Y point, so that the $X_2$-$X_3$ gap closes and reopens, but the system stays in the topologically trivial state. To explore SOC effect near this phase boundary, we examine the electronic properties at $\Delta = 2t$ and $U = 3.91t$. As shown in Fig. 4(a), small gaps of $0.02t$ appear at the X and Y points for the spin-up and spin-down channels, respectively. When SOC is introduced, these gaps first decrease with increasing $\lambda$, then close near $\lambda = 0.05t$, and reopen at larger values of $\lambda$, as depicted in Fig. 4(b). For $\lambda > 0.05t$, the SOC-induced band inversion suggests a topological transition[55]. This is confirmed by the spin-resolved anomalous Hall conductance $\sigma_{xy,\pm}$ as shown in Fig. 4(c). For $\lambda = 0.02t$, $\sigma_{xy,\pm}$ is zero at $E_F$, indicating a trivial insulating state. In contrast, for $\lambda = 0.08t$, quantized plateaus with $\sigma_{xy,\pm} = \pm e^2/h$ appear at $E_F$, confirming the emergence of the AM QSHE. The spin-resolved Berry curvature distributions in Fig. 4(d) and 4(e) show that large $\Omega(\mathbf{k})$ values are strongly localized near the X and Y points for the two spin channels. These results indicate that the AM QSHE is rather robust in a large range of U [Fig. 1(b)].

Obviously, the Lieb lattice provides an ideal platform for realizing 2D AM order and associated topological phases. We noted that the experimental realization of a freestanding AM Lieb lattice can be challenging, even though several quasi-2D altermagnets contain Lieb-lattice-based layers[40,41,43], and a number of theoretically proposed 2D altermagnets are derivatives of the Lieb lattice[33,34,35,36,37,38]. Nevertheless, appropriate substrates may stabilize a Lieb-lattice structure while preserving its magnetic and topological properties, as achieved in other systems[56,57,58]. Furthermore, analogous AM states can be also engineered in ultracold-atom systems[59] or photonic crystals[60]. Therefore, our predictions have practical significance for realizing the unusual AM QSHE state and guiding the design of spintronic and quantum devices. It is worth emphasizing that the $C_{4v}$ and mirror symmetries are fundamental symmetry requirements for establishing AM order in the Lieb lattice. Consequently, any external perturbation that breaks these symmetries, such as uniaxial strain, can suppress the AM order[61]. However, the QSHE may still be preserved under moderate uniaxial strain, provided that SOC maintains a global band gap near $E_F$. Importantly, the AM order in the Lieb lattice is intrinsically robust under equilibrium conditions, where the $C_{4v}$ and mirror symmetries are energetically favored.

In summary, we have investigated the electronic and magnetic properties of the Lieb lattice with the Hubbard model for weak and moderate electronic correlations. At an electron filling factor of $n = 2$, the Lieb lattice may host three distinct phases depending on $\Delta$ and $U$: (I) an NM metal, (II) an AM metal, and (III) an AM insulator. The AM phases appear when $U$ exceeds approximately $0.7t$. Significantly, the AM phases can be transformed into an unusual AM QSHE state by SOC in a large parameter space. This new TI state is characterized by quantized spin Hall conductance $\sigma_{xy}^S = e^2/h$ and the presence of spin-biased edge states. Our findings suggest that 2D AM order and controllable topological phases can be realized in Lieb-lattice platforms through appropriate structure engineering.

**Acknowledgements**
This work is supported by National Key Research and Development Program of China (Grant No. 2023YFA1406200), the Program for Science and Technology Innovation Team in Zhejiang (Grant No. 2021R01004), the start-up funding of Ningbo University and Yongjiang Recruitment Project (432200942).

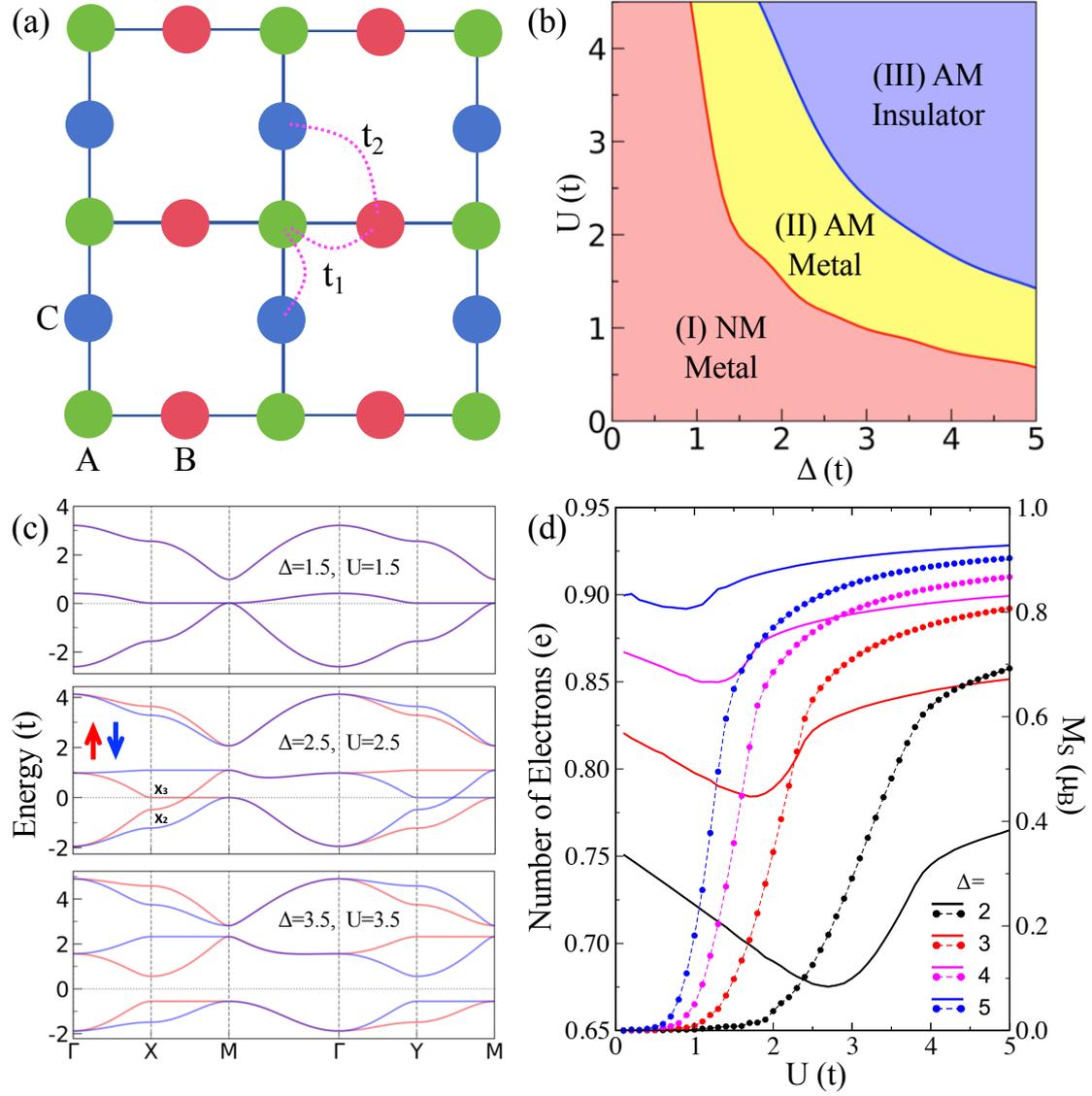

Fig. 1. (a) Structural model of two-dimensional Lieb lattice. 'A' stands for nonmagnetic sites, while 'B' and 'C' for magnetic sites with opposite spins. The parameters $t_1$ and $t_2$ indicate the hopping amplitudes between nearest and next-nearest neighbors. (b) The phase diagram of the electronic and magnetic features with respect to the values of $\Delta$ and $U$ without SOC. (c) Band structures with representative values of $\Delta$ and $U$ without SOC. The Fermi level is set to zero energy. (d) Number of electrons (solid lines) and spin moment ($M_S$) (dashed-dotted lines) on B site as a function of $U$ with selected $\Delta$.

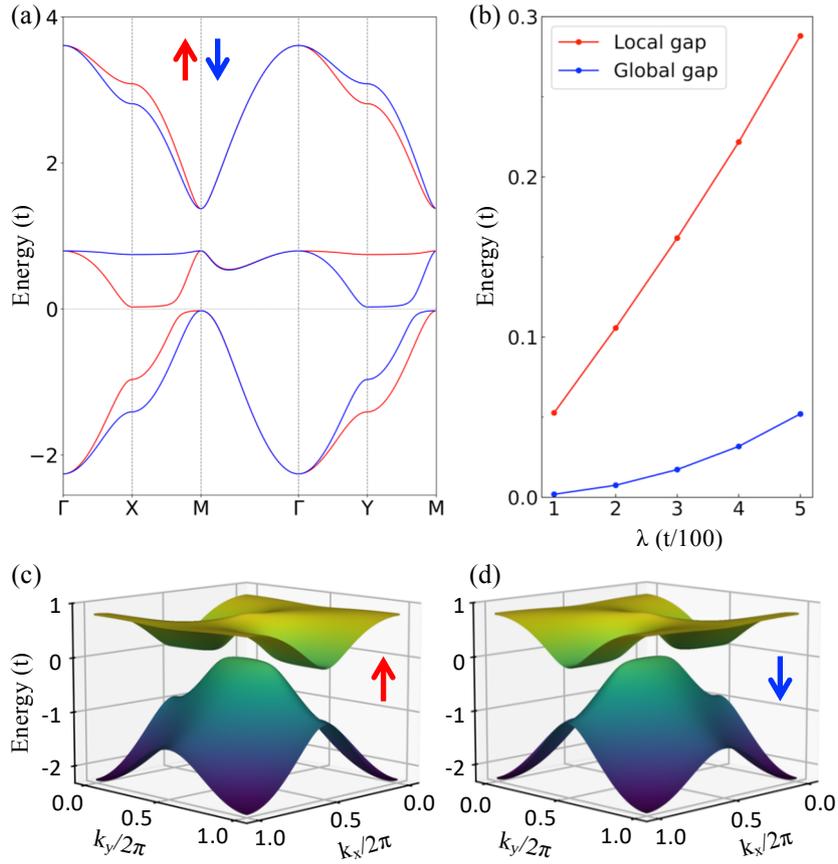

Fig. 2. (a) Band structure with $\Delta = 2$, $U = 3$, and $\lambda = 0.05$. (b) Band gaps at $E_F$ as a function of $\lambda$. (c) and (d) Three-dimensional valence band and conduction band in spin-up and spin-down channels with the same parameters as (a).

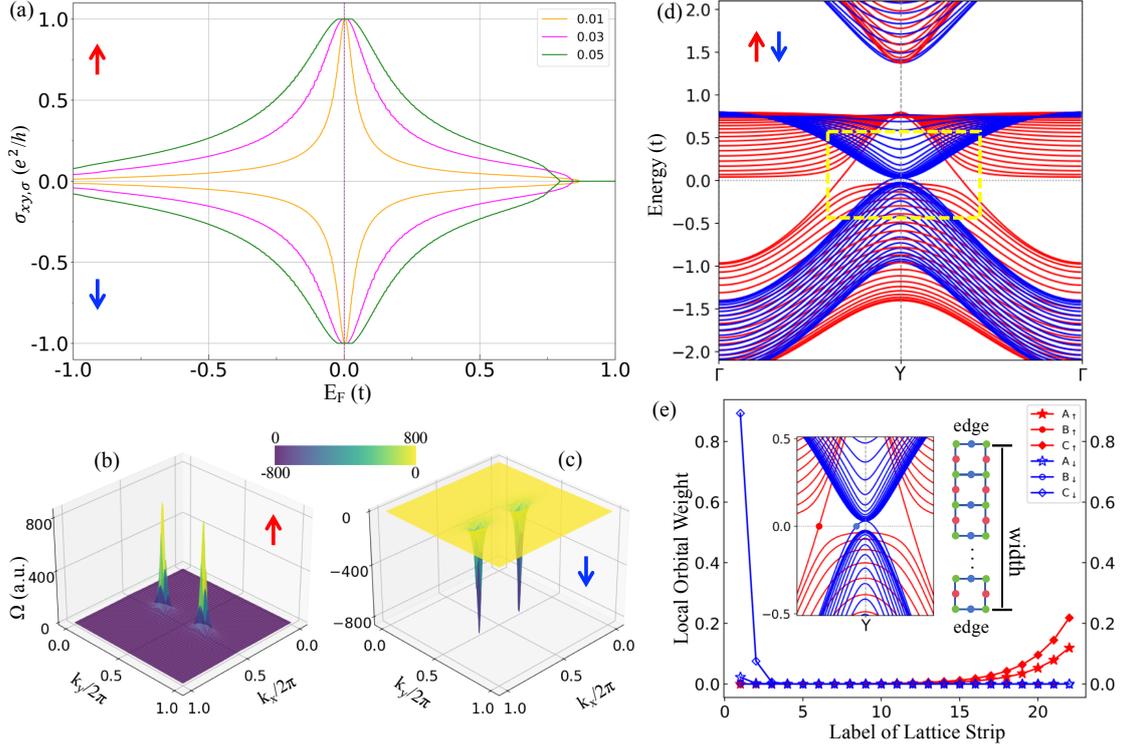

Fig. 3. (a) Spin-resolved anomalous Hall conductance ($\sigma_{xy,\pm}$) in spin-up (positive) and spin-down (negative) channels as a function of hypothetical $E_F$ with rigid band assumption. The parameters are $\Delta = 2$, $U = 3$, and $\lambda = 0.01, 0.03$, and $0.05$. (b) and (c) Distribution of Berry curvature ($\Omega$) in spin-up and spin-down channels, respectively. (d) Band structure of a nanoribbon with width of 20.5 unit cells ended by A-C chains on the edges [see the inset in (e)]. (e) Local orbital weight for the states marked by the red and blue dots at $E_F$ of the inset band structure which zooms in the region indicated by the dashed yellow rectangle in (d). The parameters for (b) to (e) are: $\Delta = 2$, $U = 3$, and $\lambda = 0.05$.

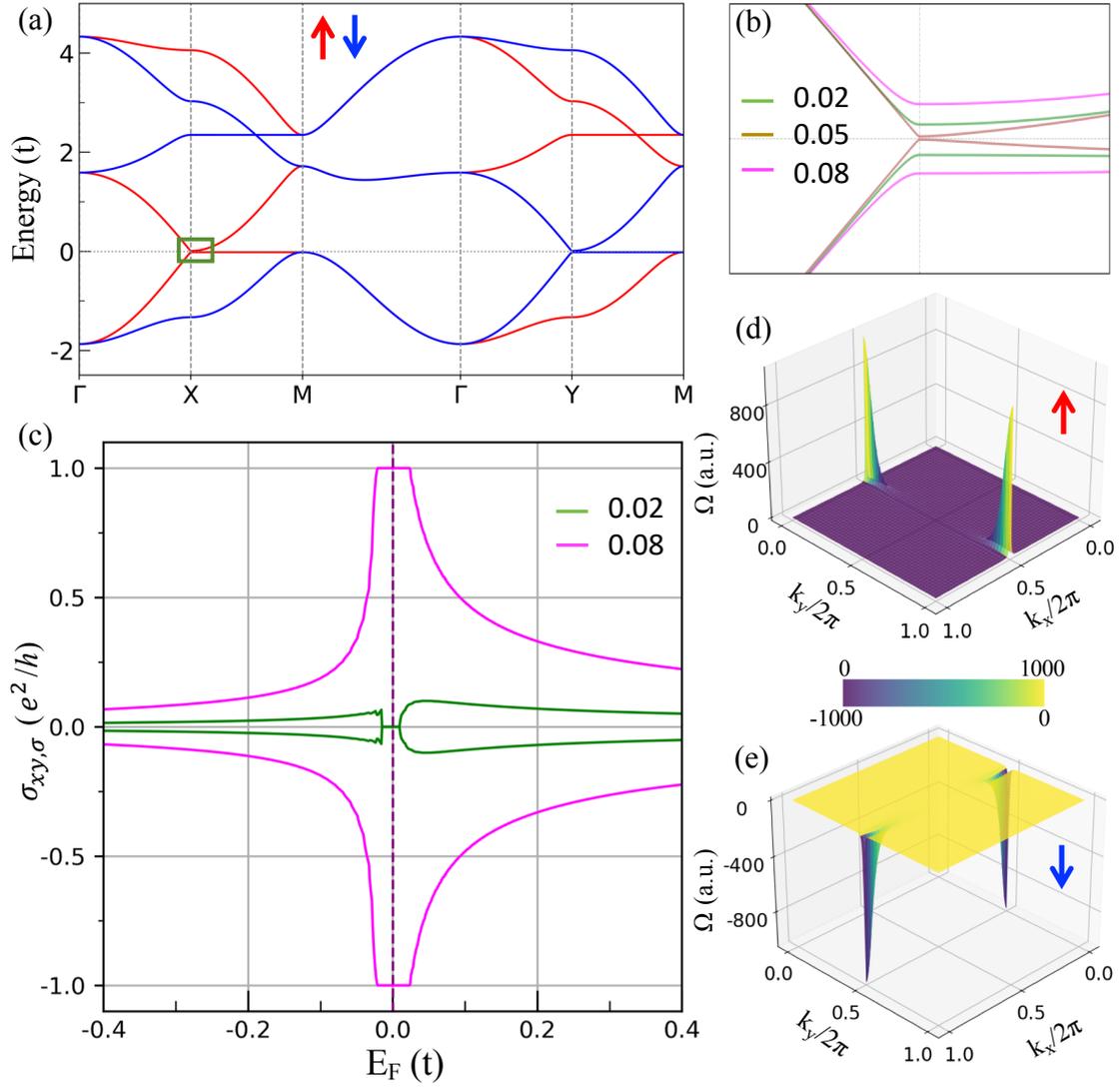

Fig.4. Electronic properties with parameters of $\Delta = 2$ and $U = 3.91$. (a) Band structure with $\lambda = 0$, i.e. without SOC. (b) Zoomed-in band structure indicated by the green rectangle in (a) with selected $\lambda$ of 0.02, 0.05 and 0.08. (c) Spin-resolved anomalous Hall conductance ($\sigma_{xy,\pm}$) in spin-up (positive) and spin-down (negative) channels as a function of hypothetical $E_F$ with rigid band assumption at $\lambda = 0.02$ and 0.08. (c) and (d) Distribution of Berry curvature ($\Omega$) in spin-up and spin-down channels, respectively, with $\lambda = 0.08$.

17